\documentclass[12pt]{article}
\usepackage{amssymb,amsmath,epsfig}

\begin{document}

\title{\bf Noether Symmetries of Some Homogeneous Universe Models in
Curvature Corrected Scalar-Tensor Gravity}

\author{M. Sharif \thanks{msharif.math@pu.edu.pk} and Saira Waheed
\thanks{smathematics@hotmail.com}\\
Department of Mathematics, University of the Punjab,\\
Quaid-e-Azam Campus, Lahore-54590, Pakistan.}

\date{}

\maketitle
\begin{abstract}
We explore Noether gauge symmetries of FRW and Bianchi I universe
models for perfect fluid in scalar-tensor gravity with extra term
$R^{-1}$ as curvature correction. Noether symmetry approach can be
used to fix the form of coupling function $\omega(\phi)$ and the
field potential $V(\phi)$. It is shown that for both models, the
Noether symmetries, the gauge function as well as the conserved
quantity, i.e., the integral of motion exist for the respective
point like Lagrangians. We determine the form of coupling function
as well as the field potential in each case. Finally, we investigate
solutions through scaling or dilatational symmetries for Bianchi I
universe model without curvature correction and discuss its
cosmological implications.
\end{abstract}

\section{Introduction}

The existence of dark energy (DE) and its role on the expansion
history of the universe has become a center of interest for the
researchers. It is a mysterious type of energy having negative
pressure that is believed to be a basic reason for the rapid
expanding behavior of the universe \cite{1,2}. For the description
of its cryptical nature, there are mainly two kinds of efforts:
modified matter approach like quintessence, Chaplygin gas, phantom,
quintom, tachyon etc. \cite{3} and the modified gravity (due to some
extra degrees of freedom) including $f(R)$ gravity, scalar-tensor
theory, $f(T)$ gravity etc. \cite{4}. Although the modified matter
approach has many novel features but this is not fully free from
ambiguities. The modified gravity approach is considered to be more
appropriate in this respect.

The dominant presence of DE in the universe leads to numerous
theoretical problems like cosmic coincidence and fine-tuning
problems \cite{5}. Scalar-tensor theories are proved to be important
efforts in the investigation of DE problem as well as various cosmic
issues like the early and late time behavior of the universe and
inflation \cite{6}. The phenomenon of cosmic acceleration can be
better described by introducing some sub-dominant terms of geometric
origin like inverse of the Ricci scalar in the Einstein-Hilbert
action. The simplest action with such modification is defined as
\cite{7}
\begin{eqnarray}\label{1}
S=\frac{1}{8\pi G}\int\sqrt{-g}(R-\frac{\mu_0^4}{R})d^4x,
\end{eqnarray}
where $R$ is the Ricci scalar, $G$ is the gravitational constant and
$\mu_0$ is an arbitrary non-zero constant. In order to be consistent
with observations and physical constraints, the action of
scalar-tensor theories, in particular, Brans-Dicke (BD) theory can
be modified in the following form \cite{8}
\begin{eqnarray}\label{2}
S=\int[\phi(R-\frac{\mu_0^4}{R})+\frac{\omega(\phi)}{\phi}g^{\mu\nu}
\nabla_\mu\phi\nabla_\nu\phi-V(\phi)+L_m]\sqrt{-g}d^4x,
\end{eqnarray}
where $\phi$ is the scalar field, $\omega(\phi)$ is the BD coupling
function, $V(\phi)$ is the field potential, $L_m$ is the matter part
of the Lagrangian and $\nabla_\mu$ indicates the covariant
derivative.

In the cosmological contexts, there are two types of Noether
symmetry techniques available in literature \cite{9}. Symmetries
which are obtained by setting the Lie derivative of the Lagrangian
to zero are called Noether symmetries. The second technique is
related with the more general symmetries known as Noether gauge
symmetries (containing the Noether symmetries as a subcase) which
involve non-zero gauge function. Noether symmetries have many
significant applications in cosmology and theoretical physics. In
particular, the existence of Noether symmetries leads to a specific
form of coupling function and the field potential in scalar-tensor
theories.

Physically, symmetries lead to the existence of conserved quantities
while on mathematical grounds, these reduce dynamics of the system
due to the presence of cyclic variables \cite{10}. Using Noether
symmetry technique, the homogeneous universe models like FRW and
Bianchi models have been discussed in $f(R)$ and scalar-tensor
gravity \cite{11}. Motavali and Golshani \cite{12} explored the form
of coupling function and the field potential for FRW universe model
using Noether symmetries. Camci and Kucukakca \cite{13} evaluated
Noether symmetries for Bianchi I, III and Kantowski-Sachs spacetimes
and discussed some field potentials. In recent papers \cite{14}, we
have explored approximate Lie and Noether symmetries of some black
holes and colliding plane waves in the framework of GR.

Kucukakca and Camci \cite{15} have obtained the function $f(R)$ and
the scale factor using Noether symmetry approach in Palatini $f(R)$
theory. Capozziello et al. \cite{16} have discussed non-static
spherically symmetric solutions in $f(R)$ gravity via Noether
symmetry analysis. Shamir et al. \cite{17} have investigated Noether
symmetries and the respective conserved quantities for FRW and
general static spherically symmetric spacetimes in $f(R)$ gravity.
Jamil et al. \cite{18} have discussed Noether gauge symmetries and
the respective conserved quantities with different forms of
potential for Bianchi I (BI) universe model in generalized
Saez-Ballester scalar-tensor gravity. Kucukakca et al. \cite{19}
have explored BI universe model through Noether symmetry analysis
with degeneracy condition of the Lagrangian and concluded that their
results are consistent with the observations. Motavali et al.
\cite{8} calculated the Noether symmetries of the Lagrangian with an
extra curvature term for FRW universe model.

Consider the point transformations (invertible transformations of
``generalized positions") that depend only upon one infinitesimal
parameter $\sigma$, i.e., $Q^i=Q^i(q^j,\sigma)$ which can generate
one-parameter Lie group \cite{14,16,18}. The vector field with
unknowns $\alpha^i$ defined by
\begin{eqnarray}\nonumber
\textbf{X}=\alpha^i(q^j)\frac{\partial}{\partial
q^i}+[\frac{d}{d\lambda}(\alpha^i(q^j))]\frac{\partial}{\partial
\dot{q}^i}
\end{eqnarray}
is said to be a Noether symmetry for the dynamics derived by the
Lagrangian if it leaves the Lagrangian invariant, that is,
$L_X\mathcal{L}=0$. In this case, the Euler-Lagrange equations and
and the constant of motion can be written as
\begin{eqnarray}\label{7}
\frac{d}{d\lambda}(\frac{\partial\mathcal{L}}{\partial\dot{q^i}})-
\frac{\partial\mathcal{L}}{\partial q^i}=0,\quad
\vartheta=\alpha^i\frac{\partial\mathcal{L}}{\partial\dot{q^i}}.
\end{eqnarray}
Noether gauge symmetries are the generalization of Noether
symmetries (as it is expected that they contain some extra
symmetries). Consider a vector field \textbf{X} as
\begin{equation*}
\textbf{X}=\tau(t,q^{i})\frac{\partial}{\partial
t}+\eta^{j}(t,q^{i})\frac{\partial}{\partial q^{j}}
\end{equation*}
and its first-order prolongation is defined as
\begin{equation}\nonumber
\textbf{X}^{[1]}=\textbf{X}+(\eta^{j}_{,t}+\eta^{j}_{,i}\dot{q}^{i}-\tau_{,t}\dot{q}^{j}
-\tau_{,i}\dot{q}^{i}\dot{q}^{j})\frac{\partial}{\partial\dot{q}^{j}}.
\end{equation}
Here $\tau$ and $\eta^j$ are the unknown functions to be determined
and $t$ is the affine parameter. The vector field $\textbf{X}$ is
said to be Noether gauge point symmetry of the Lagrangian
$\mathcal{L}(t,q^{i},\dot{q}^{i})$, if there exists a function
(known as gauge term) $G(t,q^{i})$ such that the following condition
is satisfied
\begin{equation}\label{11}
\textbf{X}^{[1]}L+(D_{t}\tau)L=D_{t}G;\quad
D_{t}=\frac{\partial}{\partial
t}+\dot{q}^{i}\frac{\partial}{\partial q^{i}}.
\end{equation}
Here $D_t$ is the total derivative operator.

There are two physical frames available in literature: Einstein and
Jordan frames which are related with each other by a conformal
transformation ($\widetilde{g}=e^{2\Omega}g$). It is argued that
both these frames are equivalent on mathematical as well as physical
grounds in the classical gravity regime where the conformal mapping
is well defined. The compatibility of the Noether symmetries and the
conformal transformations have been discussed in literature
\cite{19*}. It is proved that the Noether point symmetry if exists,
it remains preserved under the conformal transformations.

In the present paper, we evaluate Noether gauge symmetries of the
non-vacuum point like Lagrangian for FRW universe model and then
extend to locally rotationally symmetric (LRS) BI universe model,
the simplest generalization of FRW universe. The paper is designed
in the following manner. In section \textbf{2}, we evaluate Noether
gauge symmetries for FRW universe model with correction term.
Section \textbf{3} provides Noether as well as Noether gauge
symmetries for BI universe model with correction term. In section
\textbf{4}, we discuss BI solutions using scaling or dilatation
symmetries without correction term. Finally, we present an outlook
in the last section.

\section{Noether Gauge Symmetries for FRW Universe Model}

For the sake of simplicity, we take $\phi=\varphi^2$ and
$\mu_0^4=-\mu$. Thus the action for scalar-tensor gravity with extra
curvature term (\ref{2}) can be written as
\begin{eqnarray}\label{13}
S=\int[\varphi^2(R+\frac{\mu}{R})+4\omega(\varphi)g^{\mu\nu}
\nabla_\mu\varphi\nabla_\nu\varphi-V(\varphi)+L_m]\sqrt{-g}d^4x.
\end{eqnarray}
The homogeneous, non-flat FRW universe model is given by
\begin{eqnarray}\label{14}
ds^2=dt^2-a^2(t)[\frac{dr^2}{1-kr^2}+r^2(d\theta^2+\sin^2\theta
d\varphi^2)],
\end{eqnarray}
where $a(t)$ is the scale factor and $k(=0,\pm1)$ is the curvature
index. The matter part of the Lagrangian is described by the perfect
fluid
\begin{eqnarray}\label{15}
T_{\mu\nu}=(\rho+P)u_\mu u_\nu-Pg_{\mu\nu},
\end{eqnarray}
where $\rho,~P$ and $u_\mu$ denote the energy density, pressure and
four velocity, respectively. The equation of state (EoS) for perfect
fluid is $P=\epsilon\rho$, where $\epsilon$ is the EoS parameter.
The energy conservation leads to $\rho=\rho_0a^{-3(1+\epsilon)}$ and
hence the pressure becomes $P=\epsilon\rho_0a^{-3(1+\epsilon)}$.

We are interested in the Noether gauge symmetries (non-zero guage
function) of FRW model with perfect fluid matter contents. For this
purpose, the point like Lagrangian constructed by the partial
integration of the action (\ref{13}) is \cite{8}
\begin{eqnarray}\nonumber
\mathcal{L}&=&2a^3\varphi^2\mu q+6(\mu
q^2-1)(2a^2\varphi\dot{a}\dot{\varphi}+\varphi^2a\dot{a}^2)
+12\mu\varphi^2a^2q\dot{a}\dot{q}-6ka\varphi^2(\mu
q^2\\\label{16}&-&1)+a^3(4\omega(\varphi)\dot{\varphi}^2-V(\varphi))+\rho_0\epsilon
a^{-3\epsilon}.
\end{eqnarray}
In this case, the configuration space is given by $(t,a,\varphi,q)$,
consequently the Lagrangian is defined as $\mathcal{L}:
TQ\rightarrow \mathbb{R}$, where
$TQ=(t,a,\varphi,q,\dot{a},\dot{\varphi},\dot{q})$ is the respective
tangent space and $\mathbb{R}$ is the set of real numbers. The
first-order prolongation of the symmetry generator is given by
\begin{eqnarray}\nonumber
X^{[1]}&=&\tau(t,a,\varphi,q)\frac{\partial}{\partial
t}+\alpha(t,a,\varphi,q)\frac{\partial}{\partial
a}+\beta(t,a,\varphi,q)\frac{\partial}{\partial
\varphi}+\gamma(t,a,\varphi,q)\frac{\partial}{\partial
q}\\\nonumber&+&\alpha_{t}(t,a,\varphi,q)\frac{\partial}{\partial\dot{a}}
+\beta_{t}(t,a,\varphi,q)\frac{\partial}{\partial\dot{\varphi}}
+\gamma_{t}(t,a,\varphi,q)\frac{\partial}{\partial\dot{q}},
\end{eqnarray}
where $\alpha,~\beta$ and $\gamma$ are unknown functions to be
determined. Moreover,
\begin{eqnarray}\nonumber
\alpha_{t}=D_{t}\alpha-\dot{a}D_{t}\tau,\quad
\beta_{t}=D_{t}\beta-\dot{\phi}D_{t}\tau, \quad
\gamma_{t}=D_{t}\gamma-\dot{q}D_{t}\tau,
\end{eqnarray}
where
\begin{eqnarray}\nonumber
D_{t}=\frac{\partial}{\partial t}+\dot{a}\frac{\partial}{\partial
a}+\dot{\varphi}\frac{\partial}{\partial\varphi}+\dot{q}\frac{\partial}{\partial
q}.
\end{eqnarray}
Substituting these values with Eq.(\ref{16}) in Eq.(\ref{11}), we
get the following system of determining equations
\begin{eqnarray}\label{20}
&&\tau_{q}=0,\quad \tau_{a}=0,\quad \tau_{\varphi}=0,\\\nonumber
&&12(\mu q^2-1)a^2\varphi
\alpha_{t}+8a^3\omega(\varphi)(t)\beta_{t}+\tau_{\varphi}(2a^3\varphi^2\mu
q-a^3V(\varphi)\\\label{22}&&-6ka\varphi^2(\mu
q^2-1))=G_{\varphi},\\\nonumber &&12(\mu
q^2-1)a\varphi^2\alpha_{t}+12(\mu q^2-1)a^2\varphi \beta_{t}+12\mu
qa^2\varphi^2\gamma_{t}+(2a^3\varphi^2\mu
q\\\label{23}&&-a^3V(\varphi)-6ka\varphi^2(\mu
q^2-1))\tau_{a}=G_{a},\\\nonumber&& 12(\mu
q^2-1)a^2\varphi^2\alpha_{t}+(2a^3\varphi^2\mu
q-a^3V(\varphi)-6ka\varphi^2(\mu
q^2-1))\tau_{q}=G_{q},\\\label{24}\\\nonumber &&6a^2\varphi^2\mu
q\alpha-6k\varphi^2(\mu q^2-1)\alpha-3a^2\alpha
V(\varphi)+4a^3q\mu\varphi \beta-12ka\beta\varphi(\mu
q^2\\\nonumber&&-1)-a^3\beta
V'(\varphi)+2a^3\mu\varphi^2\gamma-12k\mu\varphi^2qa\gamma+(2a^3q\mu\varphi^2-6ka\varphi^2(\mu
q^2\\\label{25}&&-1)-a^3V(\varphi))\tau_{t}-3\rho_0\epsilon^2a^{-(1+3\epsilon)}\alpha=G_{t},\\\nonumber
&&24a\alpha\varphi(\mu \mu q^2-1)+12a^2\beta(\mu
q^2-1)+24a^2\varphi\mu q\gamma+12(\mu q^2-1)a^2\varphi
\alpha_{a}\\\nonumber&&+12(\mu
q^2-1)a\varphi^2\alpha_{\varphi}+12a\varphi(\mu
q^2-1)\beta_{\varphi}+12\mu
qa^2\varphi^2\gamma_{\varphi}+8a^3\omega(\varphi)\beta_{a}\\\label{26}&&-12(\mu
q^2-1)a^2\varphi\tau_{t}=0,\\\nonumber &&6(\mu
q^2-1)\alpha\varphi^2+12a\beta\varphi(\mu q^2-1)+12\mu
qa\varphi^2\gamma+12(\mu
q^2-1)a\varphi^2\alpha_{a}\\\label{27}&&+12(\mu q^2-1)a^2\varphi
\beta_{a}+12\mu qa^2\varphi^2\gamma_{a}-6a\varphi^2(\mu
q^2-1)\tau_{t}=0,\\\nonumber &&24\mu qa\varphi^2\alpha+24\mu
qa^2\varphi \beta+12\mu a^2\varphi^2\gamma+12\mu
qa^2\varphi^2\alpha_{a}+12a\varphi^2(\mu
q^2-1)\alpha_{q}\\\label{28}&&+12a^2\varphi(\mu
q^2-1)\beta_{q}+12\mu qa^2\varphi^2\gamma_{q}-12\mu
qa^2\varphi^2\tau_{t}=0,\\\nonumber
&&12a^2\omega(\varphi)\alpha+4a^3\beta\omega'(\varphi)+12a^2\varphi(\mu
q^2-1)\alpha_{\varphi}+8a^3\omega(\varphi)\beta_{\varphi}\\\label{29}&&
-4a^3\omega(\varphi)\tau_{t}=0,\\\label{30} &&12\mu
qa^2\varphi^2\alpha_{\varphi}+12(\mu q^2-1)a^2\varphi
\alpha_{q}+8a^3\omega(\varphi)\beta_{q}=0,\\\label{31}&& 12\mu
qa^2\varphi^2\alpha_{q}=0,
\end{eqnarray}
where $G=G(t,a,\varphi,q)$.

This is a system of 11 partial differential equations (PDEs) which
we solve simultaneously for the unknown functions
($\tau,~\alpha,~\beta,~\gamma,~G$). The coupling function and the
field potential both are also unknown and we specify their forms by
the existence of Noether symmetries. Integration of Eq.(\ref{31})
implies $\alpha=\alpha_1(t,a,\varphi)$. Since the above system of
PDEs is difficult to solve, therefore we take the ansatze for the
functions $\alpha_1$ and $\beta$ as
\begin{eqnarray}\label{32}
\alpha_1=\alpha_0t^{n_1}a^n\varphi^m,\quad
\beta=\beta_0(q)t^{l_1}a^l\varphi^s,
\end{eqnarray}
where $\beta_0$ is an arbitrary function and
$\alpha_0,~n,~m,~n_1,~l,~l_1,~s$ are the parameters to be
determined. Substituting these values in Eq.(\ref{30}), it follows
\begin{eqnarray}\nonumber
\beta_0(q)=-\frac{3}{4}\mu(\frac{m\alpha_0}{\omega_0})q^2+c_1, \quad
\omega(\varphi)=\omega_0\varphi^{m-s+1},\quad n=l+1, \quad n_1=l_1,
\end{eqnarray}
where $c_1$ and $\omega_0$ are constants. Equation (\ref{29}) leads
to
\begin{eqnarray}\nonumber
\tau=c_2,\quad s=m+1,\quad \omega_0=1+\frac{4c_1}{3\alpha_0},\quad
m=1.
\end{eqnarray}
Consequently, Eq.(\ref{32}) takes the form
\begin{eqnarray}\nonumber
\alpha_1=\alpha_0a^n\varphi t^{n_1},\quad
\beta=(c_1-\frac{3\alpha_0\mu}{4\omega_0}q^2)\varphi^2a^{n-1}t^{n_1}.
\end{eqnarray}
Equation (\ref{28}) implies that
\begin{eqnarray}\nonumber
\gamma(t,a,\varphi,q)=f(q)a^{n-1}t^{n_1}\varphi+\frac{g_1(t,a,\varphi)}{q},
\end{eqnarray}
where $f(q)=\frac{3\alpha_0}{2\omega_0}(\frac{\mu
q^3}{4}-\frac{q}{2})-q\alpha_0+\frac{3\mu\alpha_0q^3}{8\omega_0}-qc_1-\frac{\alpha_0nq}{2}$
and $g_1$ is an integration function. Inserting these values in
Eqs.(\ref{26}) and (\ref{27}), it follows that
\begin{eqnarray}\label{34}
\gamma=f(q)a^{n-1}t^{n_1}\varphi+\frac{g_3(t)}{aq\varphi^2},
\end{eqnarray}
where $g_3$ is an integration function.

Moreover, the following constraints should be satisfied
\begin{eqnarray}\label{35}
&&nf(q)=\frac{1-\mu q^2}{2\mu
q}[{\alpha_0(1+2n)+2n(c_1-\frac{3\alpha_0\mu}{4\omega_0}q^2)}],\\\nonumber
&&(\mu
q^2-1)[(3+n)\alpha_0+3(c_1-\frac{3\alpha_0\mu}{4\omega_0}q^2)]
+\frac{2\omega_0}{3}(n-1)(c_1-\frac{3\alpha_0\mu}{4\omega_0}q^2)\\\label{36}&&+3\mu
qf=0.
\end{eqnarray}
Integration of Eq.(\ref{24}) yields
\begin{eqnarray}\nonumber
G(t,a,\varphi,q)=12n_1\alpha_0\varphi^3t^{n_1-1}a^{n+2}(\frac{\mu
q^3}{3}-q)+h_1(t,a,\varphi),
\end{eqnarray}
where $h_1$ is an integration function. Further, Eqs.(\ref{22}) and
(\ref{23}) lead to
\begin{eqnarray}\nonumber
G=12n_1\alpha_0q(\frac{\mu
q^2}{3}-1)a^{n+2}\varphi^3t^{n_1-1}+6a^2\mu g_{3,t}+h_3(t)
\end{eqnarray}
with the constraints
\begin{eqnarray}\nonumber
&&12(\mu q^2-1)\alpha_0
n_1+8\omega_0(c_1-\frac{3\alpha_0\mu}{4\omega_0}q^2)n_1-36n_1\alpha_0q(\frac{\mu
q^2}{3}-1)=0,\\\label{37}\\\nonumber &&12(\mu
q^2-1)\alpha_0n_1+12(\mu
q^2-1)n_1(c_1-\frac{3\alpha_0\mu}{4\omega_0}q^2)+12\mu
qfn_1\\\label{38}&&=12n_1(n+2)\alpha_0q(\frac{\mu q^2}{3}-1).
\end{eqnarray}

Finally, Eq.(\ref{25}) yields
\begin{eqnarray}\nonumber
&&6\mu q\alpha_0a^{n+2}\varphi^3t^{n_1}-6k(\mu
q^2-1)\alpha_0a^n\varphi^3t^{n_1}-3V(\varphi)\alpha_0a^{n+2}\varphi
t^{n_1}+4\mu
q(c_1\\\nonumber&&-\frac{3\alpha_0\mu}{4\omega_0}q^2)a^{n+2}\varphi^3-12k(\mu
q^2-1)(c_1-\frac{3\alpha_0\mu}{4\omega_0}q^2)a^n\varphi^3t^{n_1}-a^{n+2}\varphi^2t^{n_1}
(c_1\\\nonumber&&
-\frac{3\alpha_0\mu}{4\omega_0}q^2)\frac{dV}{d\varphi}+2\varphi^3a^{n+2}f\mu
t^{n_1}+\frac{2a^2\mu g_3(t)}{q}-12kq\mu
a^{n}t^{n_1}f\varphi^3\\\nonumber&&-12k\mu
g_{3}(t)-3\rho_0\epsilon^2\alpha_0\varphi
t^{n_1}a^{-(1+3\epsilon)+n}=12n_1(n_1-1)\alpha_0q(\frac{\mu
q^2}{3}-1)\\\nonumber&&a^{n+2}\varphi^3t^{n_1-2}+6a^2\mu
g_{3,tt}+h_{3,t}.
\end{eqnarray}
This equation will be satisfied if $n_1=1$ with the following
constraints
\begin{eqnarray}\label{40}
&&-12k\mu g_3(t)+2a^2\mu g_3(t)=6a^2\mu
g_{3,tt}+h_{3,t},\\\label{41}
&&-\frac{3\alpha_0V(\varphi)}{\varphi^2}-\frac{1}{\varphi}\frac{dV}{d\varphi}=0,\\\nonumber
&&6\alpha_0q\mu-6k(\mu q^2-1)\alpha_0a^{-2}+4q\beta_0\mu-12k(\mu
q^2-1)a^{-2}\\\label{42}&&
-12kqfa^{-2}\mu-\frac{3\rho_0\epsilon^2a^{-3(1+\epsilon)\alpha_0}}{\varphi^2}=0.
\end{eqnarray}
Integration of Eqs.(\ref{40}) and (\ref{41}) yields
\begin{eqnarray}\nonumber
&&g_{3}(t)=c_{3}\exp(\sqrt{\frac{\mu}{3}}t)+c_4\exp(-\sqrt{\frac{\mu}{3}}t),\\\nonumber
&&h_3(t)=-12k\mu\sqrt{\frac{\mu}{3}}[c_{3}\exp(\sqrt{\frac{\mu}{3}}t)
+c_4\exp(-\sqrt{\frac{\mu}{3}}t)]+c_5,\\\nonumber
&&V(\varphi)=c_6\varphi^{3\alpha_0},
\end{eqnarray}
where $c_3,~c_4,~c_5$ and $c_6$ are constants of integration.

Now we can discuss Eq.(\ref{42}) for the following two cases, i.e.,
when $\epsilon=0$ or $\alpha_0=0$. If $\epsilon=0$, then pressure
becomes zero and matter distribution will be the dust dominated
fluid. Moreover, Eq.(\ref{42}) leads to the following constraints
\begin{eqnarray}\nonumber
&&6q\alpha_0\mu+4q\beta_0\mu+2f\mu=0,\\\label{40*} &&-6k(\mu
q^2-1)\alpha_0-12k(\mu q^2-1)-12kfq\mu=0.
\end{eqnarray}
In this case, the solution turns out to be
\begin{eqnarray}\nonumber
&&\alpha=\alpha_1=\alpha_0a^n\varphi t,\quad
\beta=(c_1-\frac{3\alpha_0\mu
q^2}{4\omega_0})\varphi^2a^{n-1}t,\quad \tau=c_2,\quad
V=c_6\varphi^{3\alpha_0},\\\nonumber
&&\gamma=(\frac{3\alpha_0}{2\omega_0}(\frac{\mu
q^3}{4}-\frac{q}{2})-q\alpha_0+\frac{3\mu\alpha_0q^3}{8\omega_0}
-qc_1-\frac{\alpha_0nq}{2})a^{n-1}t\varphi\\\nonumber
&&+\frac{c_{3}\exp(\sqrt{\frac{\mu}{3}}t)
+c_4\exp(-\sqrt{\frac{\mu}{3}}t)}{aq\varphi^2},\\\nonumber
&&G=12q\alpha_0(\frac{\mu
q^2}{3}-1)a^{n+2}\varphi^3+\frac{6a^2\mu^{3/2}}{\sqrt{3}}[c_{3}\exp(\sqrt{\frac{\mu}{3}}t)
-c_4\exp(-\sqrt{\frac{\mu}{3}}t)]\\\nonumber
&&-12k\mu\sqrt{\frac{\mu}{3}}[c_{3}\exp(\sqrt{\frac{\mu}{3}}t)
+c_4\exp(-\sqrt{\frac{\mu}{3}}t)]+c_5.
\end{eqnarray}
Consequently, the symmetry generator is
\begin{eqnarray}\nonumber
\textbf{X}&=&c_2\frac{\partial}{\partial t}+\alpha_0a^n\varphi
t\frac{\partial}{\partial
a}+(c_1-\frac{3\alpha_0\mu}{4\omega_0}q^2)\varphi^2a^{n-1}t\frac{\partial}{\partial\varphi}+
(f(q)a^{n-1}t\varphi\\\nonumber
&+&\frac{c_{3}\exp(\sqrt{\frac{\mu}{3}}t)+c_4\exp(-\sqrt{\frac{\mu}{3}}t)}
{aq\varphi^2})\frac{\partial}{\partial q}.
\end{eqnarray}
The corresponding conserved quantity becomes
\begin{eqnarray}\nonumber
I&=&c_2(2a^3\varphi^2\mu q+6(\mu
q^2-1)(2a^2\varphi\dot{a}\dot{\varphi}+\varphi^2a\dot{a}^2)
+12\mu\varphi^2a^2q\dot{a}\dot{q}-6ka\varphi^2(\mu q^2\\\nonumber
&&-1)+a^3(4(1+\frac{4c_1}{3\alpha_0})\dot{\varphi}^2-c_6\varphi^{3\alpha_0})
+(\alpha_0a^n\varphi t-c_2\dot{a})(6(\mu
q^2-1)(2a^2\varphi\dot{\varphi}\\\nonumber
&&+2a\varphi^2\dot{a})+12\mu
qa^2\varphi^2\dot{q})+((c_1-\frac{3\alpha_0\mu}{4\omega_0}q^2)\varphi^2a^{n-1}t-c_2\dot{\varphi})(12(\mu
q^2-1)a^2 \\\nonumber
&&\varphi\dot{a}+8a^3\dot{\varphi}\omega_0)+(f(q)a^{n-1}t\varphi
+\frac{1}{aq\varphi^2}(c_{3}\exp(\sqrt{\frac{\mu}{3}}t)
+c_4\exp(-\sqrt{\frac{\mu}{3}}t))\\\nonumber
&&-c_2\dot{q})(12\mu\varphi^2a^2q\dot{a})-12q\alpha_0(\frac{\mu
q^2}{3}-1)a^{n+2}\varphi^3-\frac{6a^2\mu^{3/2}}{\sqrt{3}}[c_{3}\exp(\sqrt{\frac{\mu}{3}}t)
\\\nonumber
&&-c_4\exp(-\sqrt{\frac{\mu}{3}}t)]+12k\mu\sqrt{\frac{\mu}{3}}[c_{3}\exp(\sqrt{\frac{\mu}{3}}t)
+c_4\exp(-\sqrt{\frac{\mu}{3}}t)]-c_5.
\end{eqnarray}

For the flat universe, the constraint (\ref{36}) restricts $q$ to be
constant say $q_0$ as follows
\begin{eqnarray}\nonumber
q^2=q_0^2=\frac{(3+n)\alpha_0+c_1-2/3(n-1)\omega_0c_1}{(1-3n)\mu}.
\end{eqnarray}
Equations (\ref{35}), (\ref{37}), (\ref{38}) and (\ref{40*}) are
four constraints that can be used to restrict the parameters
$\mu,~\alpha_0,~c_1$ and $n$. Notice that $q=1/R$, where $R$ is the
Ricci scalar, which turns out to be constant, i.e., $R=1/q_0$. This
is in agreement with Noether theorem according to which, when a
cyclic variable is identified, Noether symmetry appears and, in the
present case, the combination $R=1/q_0$ is constant which
corresponds to constant scalar curvature solution. Its physical
meaning is that the Noether symmetry generator exists for the
solutions with constant curvature like de Sitter solutions.

It is found that Noether symmetry generator exists for $\epsilon=0$
and the respective gauge function turns out to be a dynamical
quantity. Moreover, the potential is a dynamical quantity given by a
power law form while the BD coupling is a constant quantity. The
behavior of the field potential depends upon the constant $\alpha_0$
(for $\alpha_0>0$, the field potential behaves as a positive power
law while $\alpha_0<0$ leads to inverse power law potential). Such
field potentials have been used to discuss many cosmological issues
in literature \cite{21}. The existence of Noether gauge symmetries
yields the conserved quantity, i.e., Noether charge exists which can
be used to reduce the complexity of the Euler-Lagrange equations.

For the second case ($\alpha_0=0$), the field potential turns out to
be constant, i.e., $V_0=c_6$ and $\alpha=0$, also, $\omega$ is
diverging, i.e., $\omega\rightarrow\infty$, hence we neglect this
choice.

\section{Noether Gauge Symmetries for LRS Bianchi I Universe Model}

Here, we calculate the Noether and Noether gauge symmetries of the
 LRS BI spacetime. The LRS BI universe with scale factors $A$ and $B$
is defined by the line element \cite{20}
\begin{eqnarray}\label{60}
ds^2=dt^2-A^2(t)dx^2-B^2(t)(dy^2+dz^2).
\end{eqnarray}
The dynamical constraint evaluated in terms of the Ricci scalar
follows
\begin{eqnarray}\nonumber
R-2[\frac{\ddot{A}}{A}+2\frac{\ddot{B}}{B}+\frac{\dot{B}^2}{B^2}
+2\frac{\dot{A}\dot{B}}{AB}]=0.
\end{eqnarray}
Using the Lagrange multiplier approach, the action can be written as
\begin{eqnarray}\nonumber
S&=&\int[\varphi^2(R+\frac{\mu}{R})+4\dot{\varphi}^2\omega(\varphi)-V(\varphi)+\chi(R
-2\frac{\dot{A}}{A}-4\frac{\ddot{B}}{B}-2\frac{\dot{B}^2}{B^2}\\\label{61}
&-&4\frac{\dot{A}\dot{B}}{AB})+L_m](AB^2)d^4x.
\end{eqnarray}
Here $\chi$ is the Lagrange multiplier parameter. Varying this
action with respect to the Ricci scalar, the parameter $\chi$ turns
out to be
\begin{equation*}
\chi=\varphi^2(\mu R^{-2}-1).
\end{equation*}

The matter part of the Lagrangian is described by the perfect fluid
(as defined in the previous section). We consider the matter
dominated universe for which the matter part of the Lagrangian is
given by $\mathcal{L}_m=\rho_0(AB^2)^{-1}$. By substituting the
respective values in the action (\ref{61}), it follows
\begin{eqnarray}\nonumber
S&=&\int[2q\mu
AB^2\varphi^2+4AB^2\omega\dot{\varphi}^2-AB^2V-\varphi^2(\mu
q^2-1)(2A\dot{B}^2+4B\dot{A}\dot{B})\\\nonumber&-&\varphi^2(\mu
q^2-1)(2\ddot{A}B^2+4AB\ddot{B})+\rho_0]dt
\end{eqnarray}
The partial integration of this equation provides the canonical
point like form of the Lagrangian as
\begin{eqnarray}\nonumber
\mathcal{L}&=&2\mu
qAB^2\varphi^2+4AB^2\omega(\varphi)\dot{\varphi}^2-AB^2V+2\varphi^2(\mu
q^2-1)A\dot{B}^2+4B^2\varphi(\mu q^2\\\nonumber
&-&1)\dot{A}\dot{\varphi}+4\mu
qB^2\varphi^2\dot{q}\dot{A}+4B\varphi^2(\mu
q^2-1)\dot{A}\dot{B}+8AB\varphi(\mu
q^2-1)\dot{\varphi}\dot{B}\\\label{62}
&+&8AB\mu\varphi^2q\dot{B}\dot{q}+\rho_0.
\end{eqnarray}

The Euler-Lagrange equations (\ref{7}) for this Lagrangian become
\begin{eqnarray}\nonumber &&8(\mu
q^2-1)B\varphi\dot{B}\dot{\varphi}+4(\mu
q^2-1)B^2\dot{\varphi}^2+8\mu\varphi qB^2\dot{\varphi}\dot{q}+4(\mu
q^2-1)\varphi B^2\ddot{\varphi}\\\nonumber &&+8\mu
qB\varphi^2\dot{q}+8\mu q\varphi
B^2\dot{q}\dot{\varphi}+4\mu\varphi^2B^2\dot{q}^2+4\mu
q\varphi^2B^2\ddot{q}+2(\mu
q^2-1)\varphi^2\dot{B}^2\\\label{63}&&+4(\mu
q^2-1)B\varphi^2\ddot{B}-2\mu
qB^2\varphi^2-4B^2\omega(\varphi)\dot{\varphi}^2+B^2V(\varphi)=0,
\end{eqnarray}
\begin{eqnarray}\nonumber &&4(\mu q^2-1)\varphi^2A\ddot{B}+4(\mu
q^2-1)\varphi^2B\ddot{A}+8(\mu q^2-1)\varphi AB\ddot{\varphi}+8\mu
AB\varphi^2q\ddot{q}\\\nonumber&&+4(\mu
q^2-1)\varphi^2\dot{A}\dot{B}+8(\mu
q^2-1)B\varphi\dot{A}\dot{\varphi}+8\mu\varphi^2Bq\dot{A}\dot{q}+8(\mu
q^2-1)\varphi A\dot{B}\dot{\varphi}\\
\nonumber&&+8\mu\varphi^2Bq\dot{A}\dot{q}+8(\mu q^2-1)\varphi
A\dot{B}\dot{\varphi}+8\mu Aq\varphi^2\dot{q}\dot{B}+32\mu\varphi
qAB\dot{q}\dot{\varphi}+[8(\mu q^2\\\label{64}&&
-1)AB-8AB\omega(\varphi)]\dot{\varphi}^2-4q\mu
AB\varphi^2+2ABV(\varphi)=0,\\\nonumber
&&4AB^2\dot{\varphi}^2\frac{d\omega}{d\varphi}+4A\varphi(\mu
q^2-1)\dot{B}^2+AB^2\frac{dV}{d\varphi}+8AB\varphi(\mu
q^2-1)\ddot{B}+4B^2\varphi\\\nonumber&&\times(\mu
q^2-1)\ddot{A}+8AB^2\omega(\varphi)\ddot{\varphi}+8B\varphi(\mu
q^2-1)\dot{A}\dot{B}+8B^2\dot{A}\dot{\varphi}\omega(\varphi)+16A\\\label{65}&&\times
B\dot{\varphi}\dot{B}\omega(\varphi)-4\mu qAB^2\varphi=0.
\end{eqnarray}
These equations exhibit dynamics of the spatial components of the
Einstein field equations as well as scalar wave equation for BI
universe model. Another constraint on the variables can be
determined from the Ricci scalar given by
\begin{eqnarray}\nonumber
\frac{1}{q}=2[\frac{\ddot{A}}{A}+2\frac{\ddot{B}}{B}+\frac{\dot{B}^2}{B^2}
+2\frac{\dot{A}}{A}\frac{\dot{B}}{B}].
\end{eqnarray}
The energy function related with the Lagrangian $\mathcal{L}$ is
defined as \cite{8}
\begin{eqnarray}\nonumber
E_{\mathcal{L}}&=&\dot{A}\frac{\partial\mathcal{L}}{\partial\dot{A}}
+\dot{B}\frac{\partial\mathcal{L}}{\partial\dot{B}}+\dot{q}\frac{\partial\mathcal{L}}{\partial\dot{q}}
+\dot{\varphi}\frac{\partial\mathcal{L}}{\partial\dot{\varphi}}-\mathcal{L}
\end{eqnarray}
Inserting the respective values in this energy function and after
simplification, it can be written as
\begin{eqnarray}\nonumber
&&\frac{(\mu
q^2-1)\varphi^2}{6}\frac{\dot{B}^2}{B^2}+\omega(\varphi)\dot{\varphi}^2+\frac{(\mu
q^2-1)\varphi\dot{\varphi}}{3}\frac{\dot{A}}{A}+\frac{2(\mu
q^2-1)\varphi\dot{\varphi}}{3}\frac{\dot{B}}{B}\\\nonumber
&&+\frac{(\mu
q^2-1)\varphi^2}{3}\frac{\dot{A}}{A}\frac{\dot{B}}{B}+\frac{\mu\varphi^2q\dot{q}}{3}\frac{\dot{A}}{A}
+\frac{2\mu\varphi^2q\dot{q}}{3}\frac{\dot{B}}{B}-\frac{q\varphi^2\mu}{6}
+\frac{V(\varphi)}{12}-\frac{\rho_0}{12}=0.\\\label{a}
\end{eqnarray}
This provides the amount of energy in the dynamical system and
corresponds to time-time component of the field equations.
Consequently, Eqs.(\ref{63})-(\ref{a}) yield the complete set of the
field equations for BI universe.

Now we check the existence of both Noether and Noether gauge
symmetries of point like Lagrangian (\ref{62}). Here, the
configuration space for the Lagrangian is defined as
$(t,A,B,\varphi,q)$ and the respective tangent space is
$(t,A,B,\varphi,q,\dot{A}, \dot{B},\dot{\phi},\dot{q})$. The
first-order prolonged symmetry generator is defined by
\begin{eqnarray}\nonumber
\textbf{X}^{[1]}&=&\tau\frac{\partial}{\partial
t}+\alpha\frac{\partial}{\partial A}+\beta\frac{\partial}{\partial
B}+\gamma\frac{\partial}{\partial\varphi}+\delta\frac{\partial}{\partial
q}+\alpha_{t}\frac{\partial}{\partial\dot{A}}
+\beta_{t}\frac{\partial}{\partial\dot{B}}+\gamma_{t}\frac{\partial}{\partial\dot{\varphi}}
+\delta_{t}\frac{\partial}{\partial\dot{q}},\\\nonumber
\end{eqnarray}
where $\tau,~\alpha,~\beta,~\gamma$ and $\delta$ are unknown
functions to be determined. Moreover,
\begin{eqnarray}\nonumber
&&\alpha_{t}=D_{t}\alpha-\dot{A}D_{t}\tau,\quad
\beta_{t}=D_{t}\beta-\dot{B}D_{t}\tau, \quad
\gamma_{t}=D_{t}\gamma-\dot{\varphi}D_{t}\tau,\\\nonumber
&&\delta_{t}=D_{t}\delta-\dot{q}D_{t}\tau.
\end{eqnarray}
In this configuration, the total derivative operator $D_{t}$ is
\begin{eqnarray}\nonumber
D_{t}=\frac{\partial}{\partial t}+\dot{A}\frac{\partial}{\partial
A}+\dot{B}\frac{\partial}{\partial
B}+\dot{\varphi}\frac{\partial}{\partial
\varphi}+\dot{q}\frac{\partial}{\partial q}.
\end{eqnarray}
Using all these values in Eq.(\ref{11}), the system of determining
equations will beome
\begin{eqnarray}\label{100}
&&\tau_{q}=0,\quad \tau_{\varphi}=0,\quad \tau_{A}=0,\quad
\tau_{B}=0,\\\nonumber
&&B^2\alpha(2q\mu\varphi^2-V(\varphi))+AB\beta(4q\mu\varphi^2
-2V(\varphi))+AB^2\gamma(4q\mu\varphi-V'(\varphi))\\\label{101}
&&+2AB^2\mu\varphi^2\delta-\epsilon^2\rho_0(AB^2)^{-(1+\epsilon)}
(B^2\alpha+2AB\beta)=G_{t},\\\label{102}
&&4B^2q\varphi^2\mu\alpha_{t}+8ABq\mu\varphi^2\beta_{t}=G_{q},\\\nonumber
&&4\varphi^2(\mu q^2-1)B\alpha_{t}+4(\mu
q^2-1)A\varphi^2\beta_{t}+8AB\varphi(\mu
q^2-1)\gamma_{t}\\\label{103} &&+8AB\varphi^2\mu
q\delta_{t}=G_{B},\\\label{104} &&4\varphi^2(\mu
q^2-1)B\beta_{t}+4\mu B^2\varphi^2q\delta_{t}+4(\mu q^2-1)\varphi
B^2\gamma_{t}=G_{A},\\\label{105} &&8AB\varphi(\mu
q^2-1)+8AB^2\omega(\varphi)\gamma_{t}+4(\mu q^2-1)\varphi
B^2\alpha_{t}=G_{\varphi},\\\label{106}&&4\varphi^2(\mu
q^2-1)B\beta_{A}+4\mu B^2\varphi^2q\delta_{A}+4(\mu q^2-1)\varphi
B^2\gamma_{A}=0,\\\label{107}
&&4B^2\mu\varphi^2q\alpha_{q}+8AB\phi^2\mu q\beta_{q}=0,\\\nonumber
&&4B^2\alpha\omega(\varphi)+8AB\omega(\varphi)\beta+4AB^2\omega'(\varphi)\gamma+4(\mu
q^2-1)\varphi B^2\alpha_{\varphi}\\\label{108}&&+8AB(\mu
q^2-1)\varphi\beta_{\varphi}+8AB^2\omega(\varphi)\gamma_{\varphi}
-8AB^2\omega(\varphi)\tau_{t}=0,\\\nonumber &&2\alpha\varphi^2(\mu
q^2-1)+4A\varphi\gamma(\mu
q^2-1)+4q\mu\varphi^2A\delta+4\varphi^2(\mu q^2-1)B\alpha_{B}
\end{eqnarray}
\begin{eqnarray}\nonumber
&&+4\varphi^2(\mu q^2-1)A\beta_{B}-4\varphi^2(\mu
q^2-1)A\tau_{t}+8ABq\mu\varphi^2\delta_{B}\\\label{109}&&+8AB(\mu
q^2-1)\varphi\gamma_{B}=0,\\\nonumber &&4\varphi^2(\mu
q^2-1)\beta+8\varphi(\mu q^2-1)B\gamma+8\mu
qB\varphi^2\delta+4\varphi^2(\mu q^2-1)B\alpha_{A}\\\nonumber
&&-4\varphi^2(\mu q^2-1)B\tau_{t}+4\varphi^2(\mu
q^2-1)A\beta_{A}+4\varphi^2(\mu q^2-1)B\beta_{B}-4\varphi^2
\\\nonumber &&\times(\mu
q^2-1)B\tau_{t}+8ABq\varphi^2\mu\delta_{A}+4B^2q\mu\varphi^2\delta_{B}+8AB(\mu
q^2-1)\varphi\gamma_{A}\\\label{110}&&+4(\mu
q^2-1)B^2\varphi\gamma_{B}=0,\\\nonumber
&&8B\mu\varphi^2q\beta+8qB\mu\varphi\gamma+4B^2\mu\varphi^2\delta
+4B^2\mu\varphi^2q\alpha_{A}
-8B^2\mu\varphi^2q\tau_{t}+8AB\\\nonumber
&&\times\varphi^2q\mu\beta_{A}+4B\varphi^2(\mu
q^2-1)\beta_{q}+4B^2\mu\varphi^2q\delta_{q}+4(\mu
q^2-1)B^2\varphi\gamma_{q}=0,\\\label{111}&&8B\varphi(\mu
q^2-1)\beta+4(\mu q^2-1)B^2\gamma+8qB^2\mu\varphi\delta+4(\mu
q^2-1)B^2\varphi\alpha_{A}\\\nonumber&&-8(\mu
q^2-1)B^2\varphi\tau_{t}+8AB\varphi(\mu
q^2-1)\beta_{A}+4\varphi^2(\mu
q^2-1)B\beta_{\varphi}+4B^2q\mu\\\label{112}
&&\varphi^2\delta_{\varphi}+8AB^2\omega(\varphi)\gamma_{A}+4B^2(\mu
q^2-1)\varphi\gamma_{\varphi}=0,\\\nonumber
&&8Bq\mu\varphi^2\alpha+8Aq\mu\varphi^2\beta+16qAB\mu\varphi\gamma
+8AB\mu\varphi^2\delta+4B^2q\mu\varphi^2\alpha_{B}\\\nonumber
&&+4B\varphi^2(\mu
q^2-1)\alpha_{q}+8ABq\mu\varphi^2\beta_{B}+4\varphi^2A(\mu
q^2-1)\beta_{q}-16ABq\mu\varphi^2\tau_{t}\\
&&+8ABq\mu\varphi^2\delta_{q}+8AB\varphi(\mu
q^2-1)\gamma_{q}=0,\\\nonumber &&8B\varphi(\mu
q^2-1)\alpha+8A\varphi(\mu q^2-1)\beta+8AB(\mu
q^2-1)\gamma+16ABq\mu\varphi\delta\\\nonumber&&+4(\mu q^2
-1)B^2\varphi\alpha_{B}+4(\mu
q^2-1)B\varphi^2\alpha_{\varphi}+8AB(\mu
q^2-1)\varphi\beta_{B}+4A\\\nonumber&&\times(\mu
q^2-1)\varphi^2\beta_{\varphi}-16AB(\mu
q^2-1)\varphi\tau_{t}+8ABq\mu\varphi^2\delta_{\varphi}+8AB^2\omega(\varphi)\gamma_{B}
\\\label{114}&&+8AB(\mu
q^2-1)\varphi\gamma_{\varphi}=0,\\\nonumber
&&4B^2q\mu\varphi^2\alpha_{\varphi}+4(\mu
q^2-1)B^2\varphi\alpha_{q}+8ABq\mu\varphi^2\beta_{\varphi}+8AB(\mu
q^2-1)\varphi\beta_{q}\\\label{115}
&&+8AB^2\omega(\varphi)\gamma_{q}=0.
\end{eqnarray}

Firstly, we calculate the Noether symmetries of Lagrangian that
correspond to $\mathcal{L}_{X}L=0$ and can be found by the system of
determining equations (\ref{100})-(\ref{115}) with $G=0$ and
$\tau=0$. In this case, all the functions $\alpha,~\beta,~\gamma$
and $\delta$ are independent of time. Integration of Eq.(\ref{100})
yields $\tau=\tau(t)$. For the sake of simplicity, we take the
ansatz for unknowns $\alpha,~\beta$ and $\gamma$ as
\begin{eqnarray}\label{116}
\alpha=A^aB^b\varphi^cq_0(q),\quad
\beta=A^fB^g\varphi^hq_1(q),\quad\gamma=A^mB^n\varphi^pq_2(q).
\end{eqnarray}
Here $a,b,c,f,g,h,m,n$ and $p$ are parameters to be determined,
while $q_0,~q_1$ and $q_2$ are unknown functions of variable $q$. We
would like to find the functions $\delta,~V$ and $\omega$ by
requiring the existence of Noether symmetries. From Eq.(\ref{107}),
it follows that
\begin{eqnarray}\nonumber
a=f+1,\quad g=b+1,\quad c=h,\quad q_0=-2q_1,
\end{eqnarray}
and hence
\begin{eqnarray}\label{117}
\alpha=-2A^{f+1}B^{g-1}\varphi^hq_1(q),\quad
\beta=A^fB^g\varphi^hq_1(q),\quad \gamma=A^mB^n\varphi^pq_2(q).
\end{eqnarray}
Equations (\ref{108}) and (\ref{115}) imply that
$\omega(\varphi)=\frac{c_2}{\varphi^{2p}}$ and $q_2=c_1$,
respectively, where $c_1$ and $c_2$ are integration constants.
Equation (\ref{106}) yields
\begin{eqnarray}\nonumber
\delta=\frac{(1-\mu q^2)}{\mu
q}[A^fB^{g-1}\varphi^hq_1+c_1A^mB^n\varphi^{p-1}]+h_1(B,q,\varphi),
\end{eqnarray}
where $h_1$ is an integration function. Further, Eq.(\ref{109})
leads to $h_1=\frac{h_2(q,\varphi)}{\sqrt{B}}$ and $g=2/3$, hence
\begin{eqnarray}\nonumber
&&\alpha=-2q_1A^{f+1}B^{-1/3}\varphi^{h},\quad\beta=q_1A^fB^{2/3}\varphi^h,
\quad\gamma=A^mB^n\varphi^pc_1,\\\nonumber&&\delta=\frac{1-\mu
q^2}{\mu
q}[A^fB^{-1/3}\varphi^hq_1+c_1A^mB^n\varphi^{p-1}]+\frac{h_2(q,\varphi)}{\sqrt{B}}.
\end{eqnarray}
Equation (\ref{110}) implies that $h_2=0$ and $f=-2/3$. Moreover,
Eq.(\ref{112}) leads to $m=-2/3,~n=-1/3,~h=-(1+p)$ and
$q_1=\frac{2c_1c_2}{3(1-\mu q^2)}$, thus
\begin{eqnarray}\nonumber
&&\alpha=-2(\frac{2c_1c_2}{3(1-\mu
q^2)})A^{1/3}B^{-1/3}\varphi^{-(1+p)},\quad
\beta=\frac{2c_1c_2}{3(1-\mu
q^2)}A^{-2/3}B^{2/3}\varphi^{-(1+p)},\\\nonumber&&
\gamma=A^{-2/3}B^{-1/3}c_1\varphi^p,\quad\delta=\frac{(1-\mu
q^2)}{\mu q}[A^{-2/3}B^{-1/3}\varphi^{-(1+p)}\frac{2c_1c_2}{3(1-\mu
q^2)}\\\label{118}&&+c_1A^{-2/3}B^{-1/3}\varphi^{p-1}].
\end{eqnarray}
Inserting these values in Eq.(\ref{111}), we obtain either
$c_1c_2=0$ or $\mu=0$. Since $\mu\neq0$, so $c_1c_2=0$. When we take
$c_1\neq0,~c_2=0$, it follows that
\begin{eqnarray}\nonumber
&&\alpha=0,\quad \beta=0,\quad \omega=0,\quad
\gamma=c_1\varphi^pA^{-2/3}B^{-1/3},\\\label{120}
&&\delta=\frac{1-\mu q^2}{\mu q}c_1A^{-2/3}B^{-1/3}\varphi^{p-1}.
\end{eqnarray}
Equation (\ref{101}) leads to
$V(\varphi)=\frac{c_3\varphi^2}{2}+c_4$ and
$q=q_0=\frac{c_3\pm\sqrt{c_3^2-8\mu}}{4\mu}$, where $c_3$ and $c_4$
are integration constants. In this case, the symmetry generator
follows
\begin{eqnarray}\label{121}
\textbf{X}_1=\varphi^pA^{-2/3}B^{-1/3}\frac{\partial}{\partial\varphi}+\frac{1-\mu
q^2}{\mu q}A^{-2/3}B^{-1/3}\varphi^{p-1}\frac{\partial}{\partial q}
\end{eqnarray}
which yields only one symmetry and the respective constant of motion
is zero, i.e., $I_1=0$. This shows that for Noether symmetries of
the point like Lagrangian exists but there is no non-trivial
conserved quantity.

For Noether gauge symmetries, we consider the full symmetry
generator ($\tau\neq0,~G\neq0$) in which the unknown functions are
dependent on time. Proceeding in the similar way, Eqs.(\ref{100})
and (\ref{106})-(\ref{115}) lead to
\begin{eqnarray}\label{123}
\tau=c_3,\quad \alpha=0,\quad \beta=0,\quad
\gamma=c_1t^l\varphi^p,\quad \delta=\frac{1-\mu q^2}{\mu
q}c_1t^l\varphi^{p-1},\quad \omega=\frac{c_4}{\varphi^{2p}}.
\end{eqnarray}
Equations (\ref{101})-(\ref{105}) yield $l=0$ and $q=q_0$ with
\begin{eqnarray}\nonumber
V(\varphi)=\frac{1+3q_0^2\mu}{2q_0}\varphi^2, \quad G=c_5.
\end{eqnarray}
Thus there exist two symmetry generators given by
\begin{eqnarray}\nonumber
\textbf{X}_1=\frac{\partial}{\partial t},\quad
\textbf{X}_2=\varphi^p\frac{\partial}{\partial\varphi}+(\frac{1-\mu
q^2}{\mu q})\varphi^{p-1}\frac{\partial}{\partial q}.
\end{eqnarray}
The constant of motion, i.e., the integral of motion can be written
as
\begin{eqnarray}\nonumber
I=\tau\mathcal{L}+(\alpha-\dot{A}\tau)\frac{\partial\mathcal{L}}{\partial
\dot{A}}+(\beta-\dot{B}\tau)\frac{\partial\mathcal{L}}{\partial
\dot{B}}+(\gamma-\dot{\varphi}\tau)\frac{\partial\mathcal{L}}{\partial
\dot{\varphi}}+(\delta-\dot{q}\tau)\frac{\partial\mathcal{L}}{\partial
\dot{q}}-G.
\end{eqnarray}
In this case, these are given by
\begin{eqnarray}\nonumber
I_1&=&8AB^2c_4\dot{\varphi}\varphi^{-p}-\frac{c_5}{2},\\\nonumber
I_2&=&2q_0\mu
AB^2\varphi^2-4AB^2\dot{\varphi}^2c_4\varphi^{-2p}-AB^2\frac{(1+3\mu
q_0^2)}{2q_0}\varphi^2-2(\mu q_0^2-1)\varphi^2\dot{A}B^2\\\nonumber
&-&4(\mu
q_0^2-1)(B^2\varphi\dot{A}\dot{\varphi}+B\varphi^2\dot{A}\dot{B}
+2AB\varphi\dot{B}\dot{\varphi})+\rho_0-\frac{c_5}{2}.
\end{eqnarray}

It can be concluded that for the point-like Lagrangian of BI
universe model, there is only one Noether symmetry generator while
two Noether gauge symmetry generators exist. The existence of
Noether symmetries allows zero BD coupling function and the
quadratic field potential. Such field potentials have widely been
used in literature \cite{21} to discuss many cosmological problems
in the context of scalar tensor gravity. Since, $\omega=0$,
therefore these symmetries may correspond to the symmetries of
pointlike Lagrangian of BI universe in Palatini $f(R)$ gravity
\cite{22}.

It is found that the Noether charge is zero in the case of Noether
symmetries. However, the existence of Noether gauge symmetries leads
to dynamical BD coupling parameter (in the form of inverse power
law) with quadratic potential. It is observed that the behavior of
BD coupling function depends upon the parameter $p$, for $p>0$, the
BD coupling becomes divergent at $\varphi=0$ while for $p<0$, it
turns out to be zero there. Moreover, in this case, the gauge
function turns out to be constant and the Noether charge, i.e., the
conserved quantities exist. The EoS parameter for this configuration
is given by
\begin{eqnarray}\nonumber
\omega_\varphi&=&\frac{\omega_{x\varphi}+2\omega_{y\varphi}}{3}=\frac{1}{3}[3\mu
q_0\varphi^2-3/2V(\varphi)+6\omega(\varphi)\dot{\varphi}^2-6\dot{\varphi}^2(\mu
q_0^2-1)\\\nonumber &-&6\dot{\varphi}\ddot{\varphi}(\mu
q_0^2-1)-4(\mu
q_0^2-1)\varphi\dot{\varphi}(\frac{\dot{A}}{A}+2\frac{\dot{B}}{B})][\mu
q_0\varphi^2-\frac{V(\varphi)}{2}-6\omega(\varphi)\dot{\varphi}^2\\\nonumber
&-&2(\mu
q_0^2-1)\varphi(\frac{\dot{A}}{A}+2\frac{\dot{B}}{B})]^{-1},
\end{eqnarray}
where $V$ and $\omega$ are given in previous cases (found by the
Noether symmetry analysis). Following \cite{18}, we have tried to
plot this expression using Maple software but due to highly
non-linear terms present in the field equations with $A,~B$ and
$\varphi$ as unknowns, it is not possible to have the plot of this
expression (basically, Maple could not convert the expressions into
explicit first-order system of DEs).

\section{Bianchi I Solutions Using Scaling Symmetries}

In this section, we discuss BI solutions by taking $\mu=0$ in the
Lagrangian with constant BD parameter. In canonical form, the action
can be written as
\begin{equation}\label{1*}
S=\int
\sqrt{-g}[\frac{1}{8\omega}\varphi^2R-\frac{1}{2}g^{\mu\nu}\partial_\mu\varphi\partial_\nu\varphi
+V_0\varphi^2+\mathcal{L}_m]d^4x.
\end{equation}
Here $\omega$ is a constant BD parameter and the field potential is
taken to be $V=V_0\varphi^2$. Moreover, the matter distribution is
taken as the perfect fluid. The corresponding field equations are
\begin{eqnarray}\label{2*}
&&\frac{\varphi^2}{4\omega}(2\frac{\dot{A}}{A}\frac{\dot{B}}{B}
+\frac{\dot{B}}{B})+\frac{\dot{\varphi}^2}{2}
+V_0\varphi^2+\frac{1}{2\omega}(\frac{\dot{A}}{A}+2\frac{\dot{B}}{B})
\varphi\dot{\varphi}=\rho,\\\nonumber
&&-\frac{\varphi^2}{4\omega}(2\frac{\ddot{B}}{B}+\frac{\dot{B}^2}{B})
-\frac{1}{\omega}\frac{\dot{B}}{B}\dot{\varphi}\varphi
-\frac{1}{2\omega}\ddot{\varphi}\varphi
-V_0\varphi^2+(1/2-1/2\omega)\dot{\varphi}^2=P,\\\label{3*}\\\nonumber
&&-\frac{\dot{\varphi}^2}{4\omega}(\frac{\ddot{B}}{B}+\frac{\ddot{A}}{A}
+\frac{\dot{A}}{A}\frac{\dot{B}}{B})
-\frac{1}{2\omega}(\frac{\dot{A}}{A}+\frac{\dot{B}}{B})\dot{\varphi}
-\frac{1}{2\omega}\ddot{\varphi}\varphi -V_0\varphi^2\\\label{4*}&&
+(\frac{1}{2}-\frac{1}{2\omega})\dot{\varphi}^2=P,\\\label{5*}
&&\ddot{\varphi}+(\frac{\dot{A}}{A}+2\frac{\dot{B}}{B})\dot{\varphi}
+[V_0+\frac{1}{2\omega}\{\frac{\dot{A}}{A}
+2\frac{\ddot{B}}{B}+\frac{\dot{B}^2}{B^2}+2\frac{\dot{A}\dot{B}}{AB}\}]\varphi=0.
\end{eqnarray}
The equation of continuity is
\begin{eqnarray}\label{6*}
\dot{\rho}+(\frac{\dot{A}}{A}+2\frac{\dot{B}}{B})(\rho+P)=0.
\end{eqnarray}

Due to complexity of the system, we use the physical relationship
between the scale factors, i.e., $A=B^m;~m\neq1$ \cite{20}. This
condition is originated by the assumption that the ratio of shear
scalar to expansion scalar is constant. Consider the power law form
for the density $\rho=\rho_0(AB^2)^\varepsilon/3$, and hence
pressure $P=-\frac{\varepsilon+3}{3}\rho$. Here $\varepsilon$ is any
parameter that acts as an equation of state parameter and its
different values can classify different phases of the universe,
e.g., $\varepsilon=-3$ implies matter dominated era and
$\varepsilon=0$ yields dark energy dominated universe. We take
dependent variables $H_2=H_2(B)$ and $F=F(B_2)$ with $B$ as
independent variable and introduce the following notations in the
system of equations (\ref{2*})-(\ref{5*})
\begin{eqnarray}\nonumber
&&F=\frac{\dot{\varphi}}{\varphi},\quad H_2=\frac{\dot{B}}{B}, \quad
H_1=m\frac{\dot{B}}{B},\quad \dot{H_2}=BH_2H'_2,\quad
\dot{H}_1=mBH_2H'_2,\quad\\\nonumber
&&\frac{\ddot{B}}{B}=BH_2H'_2+H_2^2,\quad\frac{\ddot{\varphi}}{\varphi}=BH_2F'+F^2,
\end{eqnarray}
where prime indicates derivative with respect to $B$. This leads to
\begin{eqnarray}\label{7*}
&&H_2^2+(F^2+2V_0)\frac{2\omega}{1+2m}+\frac{2(m+2)}{2m+1}H_2F
=\frac{4\omega\rho_0B^{\varepsilon(m+2)/3}}{(1+2m)\varphi^2},
\end{eqnarray}
\begin{eqnarray}\label{8*}
&&\frac{2BH_2}{3}(F'+H'_2)+H_2^2+\frac{4H_2F}{3}+\frac{2(2-\omega)F^2}{3}
+\frac{4V_0\omega}{3}=-\frac{4\omega}{3}\frac{P}{\varphi^2},\\\nonumber
&&H_2^2+\frac{1}{(m^2+m+1)}[BH_2((m+1)H'_2+2F')+2(1+m)H_2F\\\label{9*}
&&+4V_0\omega+2(2-\omega)F^2]=-\frac{4P\omega}{(m^2+m+1)\varphi^2},\\\nonumber
&&BH_2[\frac{\omega}{m+2}F'+\frac{H'_2}{2}]+\frac{H_2^2}{2}(\frac{m^2+2m+3}{m+2})+\omega
H_2F\\\label{10*}&&+\frac{\omega}{m+2}(F^2+2V_0)=0.
\end{eqnarray}
Also, the evolution of energy density is given by
\begin{eqnarray}\nonumber
\dot{\rho}=\varepsilon\rho_0\frac{(m+2)H_2\rho}{3}.
\end{eqnarray}
In the above system of five field equations, only three equations
are independent. We shall use Eqs.(\ref{6*}), (\ref{7*}) and
(\ref{10*}) as independent equations. By taking the time derivative
of Eq.(\ref{7*}), and after some manipulation, it becomes
\begin{eqnarray}\nonumber
&&[H_2^2+\frac{m+2}{1+2m}FH_2]H'_2+\frac{F'}{1+2m}[2\omega
FH_2+(m+2)H_2^2]=\frac{\varepsilon(m+2)H_2^3}{6B}\\\nonumber
&&+\frac{FH_2^2}{B}(\frac{\varepsilon(m+2)^2}{3(1+2m)}-1)-\frac{2F^3\omega}{B(1+2m)}
+(\varepsilon\omega-6)\frac{H_2F^2(m+2)}{3B(1+2m)}\\\label{11*}
&&+\frac{2V_0\omega}{B(1+2m)}[\frac{\varepsilon(m+2)H_2}{3}-2F].
\end{eqnarray}

We can write equations for the unknowns $H_2'$ and $F'$ after some
manipulation from Eqs.(\ref{10*}) and (\ref{11*}) as follows
\begin{eqnarray}\nonumber
&&BH_2\frac{dH_2}{dB}[\frac{2(\omega-1)+m(4\omega-1)}{2(m+2)(1+2m)}]
=H_2^2[\frac{\varepsilon\omega}{6}
+\frac{m^2+2m+3}{2(1+2m)}]+F^2\omega\\\nonumber
&&\times[\frac{(\varepsilon+6)\omega-3}{3(1+2m)}]+\omega
FH_2[\frac{\varepsilon(m+2)^2+3(m-1)}{3(1+2m)}]+\frac{2V_0\omega}{3(1+2m)}
(\varepsilon\omega+3),\\\label{12*}\\\nonumber
&&BH_2(\frac{m+2}{2(1+2m)}-\frac{\omega}{m+2})\frac{dF}{dB}=H_2^2[\frac{\varepsilon(m+2)^2
+6(m^2+2m+3)}{12(m+2)}]\end{eqnarray}
\begin{eqnarray}\nonumber
&&+\frac{FH_2}{2}[\frac{\varepsilon(m+2)^2}{3(1+2m)}
-1+2\omega+\frac{m^2+2m+3}{1+2m}]+F^2[\frac{(m+2)(\varepsilon\omega-6)}{6(1+2m)}
\\\label{13*}
&&+\frac{\omega}{m+2}+\frac{\omega(m+2)}{1+2m}]+V_0\omega[\frac{6(1+2m)
+\varepsilon(m+2)^2}{3(1+2m)(m+2)}].
\end{eqnarray}
These form a closed system of equations (two equations involving two
unknowns). We adopt the analysis of classical Lie groups \cite{23}
to find solution of Eqs.(\ref{12*}) and (\ref{13*}). Since these
equations are quite similar to the equations exhibiting scaling or
dilatational symmetries, therefore we assume a vector field
\begin{eqnarray}\nonumber
\textbf{X}=B\alpha\frac{\partial}{\partial B}+\beta
F\frac{\partial}{\partial F}+\gamma H_2\frac{\partial}{\partial H_2}
\end{eqnarray}
generated by a scaling group of mappings given by
\begin{eqnarray}\nonumber
\widetilde{B}=\lambda^\alpha B,\quad
\widetilde{A}=\lambda^{m\alpha}A,\quad \widetilde{F}=\lambda^\beta
F,\quad \widetilde{H}_2=\lambda^\gamma
H_2,\quad\widetilde{H}_1=m\lambda^\gamma H_2.
\end{eqnarray}
The invariance of Eqs.(\ref{12*}) and (\ref{13*}) under the above
transformation provides two different cases:
\begin{itemize}
\item The scalar field is massive, i.e., $V_0\neq0$ which implies
that $\beta=\gamma=0$ and $\alpha=1$. This allows the form of
generator given by $\textbf{X}_1=B\frac{\partial}{\partial B}$.
\item The scalar field is massless, i.e., $V_0=0$ which yields two choices
of the parameters and consequently two symmetry generators exist.
(i) $\beta=\gamma=0$ with $\alpha=1$ implies
$\textbf{X}_1=B\frac{\partial}{\partial B}$ (ii) $\beta=\gamma=1$
and $\alpha=0$, leading to $\textbf{X}_2=H_2\frac{\partial}{\partial
H_2}+F\frac{\partial}{\partial F}$.
\end{itemize}

In the first case, there is only one symmetry generator with basis
of invariants $\{H_2,F\}$. This symmetry ensures that the solution
is in the form of invariants given by $F=F(H_2)$. In order to find
the solution, the number of known symmetries should be equal to the
order of DE. Since Eqs.(\ref{12*}) and (\ref{13*}) lead to
$\frac{dF}{dH_2}=K(H_2,F)$, i.e., first-order differential with no
more known symmetries, so the integration in quadratures will not be
possible. We construct a solution by imposing the invariance of $F$
and $H_2$, i.e., $\frac{dH_2}{dB}=0$ and $\frac{dF}{dB}=0$.
Equations (\ref{12*}) and (\ref{13*}) become quadratic equations for
$H_2$ and $F$. The roots of these equations are quite lengthy. To
get insights, we choose $m=2,~U_0=2$ and $\varepsilon=-3$ (matter
dominated phase) which yields four roots as
\begin{eqnarray}\nonumber
H_{2}&=&\pm[\omega(11264-15936\omega+28912\omega^2-4940\omega^3
-9600\omega^4-1600\omega^5)\\\nonumber
&\pm&20\omega^2(3748096-11585024\omega
+16537312\omega^2-11402336\omega^3+3126193\\\nonumber
&\times&\omega^4-557336\omega^5+197920\omega^5
+92800\omega^6+6400\omega^7)^{1/2}]^{1/2}[2000\omega^5\\\nonumber
&+&1400\omega^4-83035\omega^3+111598\omega^2-6512\omega-15488]^{-1}.
\end{eqnarray}
Likewise, there exist four roots for $F$ given by
\begin{eqnarray}\nonumber
F&=&\pm[\omega(25696-53008\omega+29310\omega^2-25600\omega^3
+4000\omega^4)\pm10\omega(3748096\\\nonumber
&-&24873728\omega+64116448\omega^2-8571569\omega^3+65210905\omega^4
-2517980\omega^5\\\nonumber
&+&8538400\omega^5-1904000\omega^6+160000\omega^7)^{1/2}]^{1/2}
[2000\omega^5+1400\omega^4-83035\\\nonumber
&\times&\omega^3+111598\omega^2-6512\omega-15488]^{-1}.
\end{eqnarray}
Consequently, the scale factor ($H_2=constant$) and scalar field
($F=constant$) turn out to be $B=B_0\exp(H_{2i}t)$ and
$\varphi=\varphi_0\exp(F_it)$, respectively. Here $H_{2i}$ and $F_i$
are the roots given above, while $B_0$ and $\varphi_0$ are present
values of the scale factor and the scalar field, respectively. The
present values can be restricted by using Eq.(\ref{7*}) as follows
\begin{equation*}
\varphi_0=\sqrt{\frac{4\omega\rho_0B_0^{-4}}{(5H_{2i}^2+2\omega
F_i^2+8\omega+8H_{2i}F_i)}}.
\end{equation*}

The plots for the scale factor and density function
($\rho(t)=\rho_0B_0^{4\epsilon}e^{4\epsilon H_{2i}t}/3$) are shown
in Figure \textbf{1}. Since the scalar field is massive, therefore
$\omega$ can take any value satisfying $\omega>-3/2$ to avoid ghost
instabilities \cite{22}. Figure \textbf{1(a)} indicates that the
scale factor increases for $(-,-)$ and $(+,+)$ roots, while
decreases to zero for $(-,+)$ and $(+,-)$ roots. Figure
\textbf{1(b)} shows that the behavior of density is exactly opposite
to that of the scale factor. This means that only for $(-,-)$ and
$(+,+)$ roots, the constructed model shows expanding behavior with
decreasing energy density which is consistent with the recent
observations. The scalar field exhibits a similar behavior as shown
in Figure \textbf{2(a)}.
\begin{figure} \centering
\epsfig{file=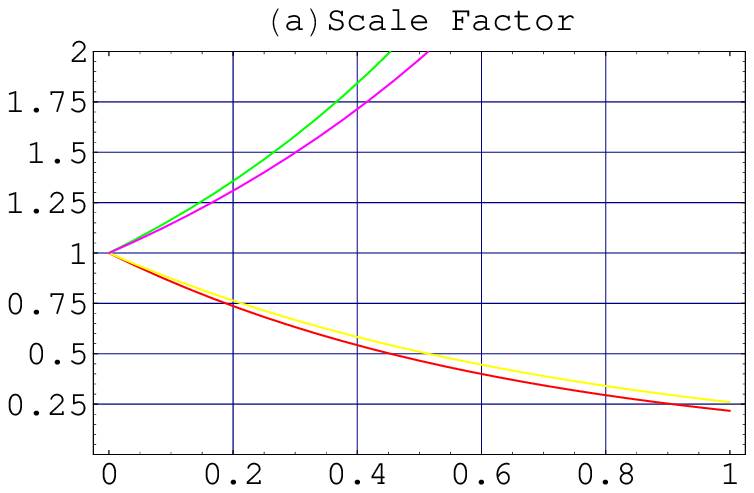,width=.45\linewidth}
\epsfig{file=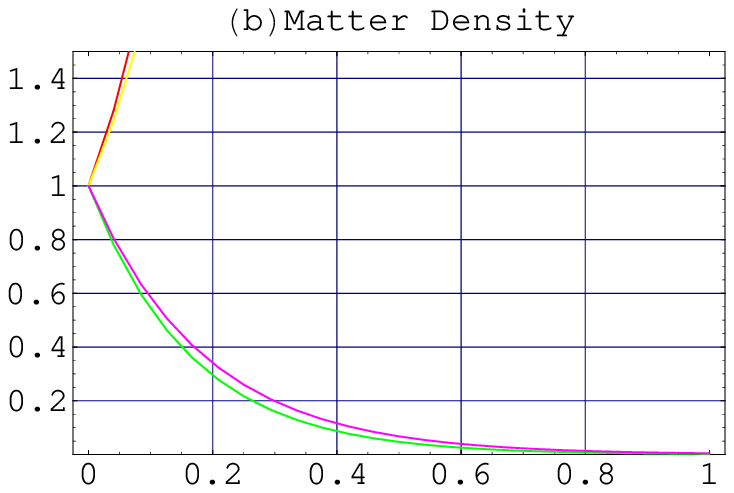,width=.45\linewidth} \caption{Plots (a) and
(b) show the scale factor and energy density versus time $t$,
respectively. Here red and green correspond to $(+,+)$ and $(-,-)$
roots with $\omega=1.5$, respectively, while yellow and purple lines
indicate $(+,-)$ and $(-,+)$ roots with $\omega=1.2$, respectively.}
\end{figure}
\begin{figure} \centering
\epsfig{file=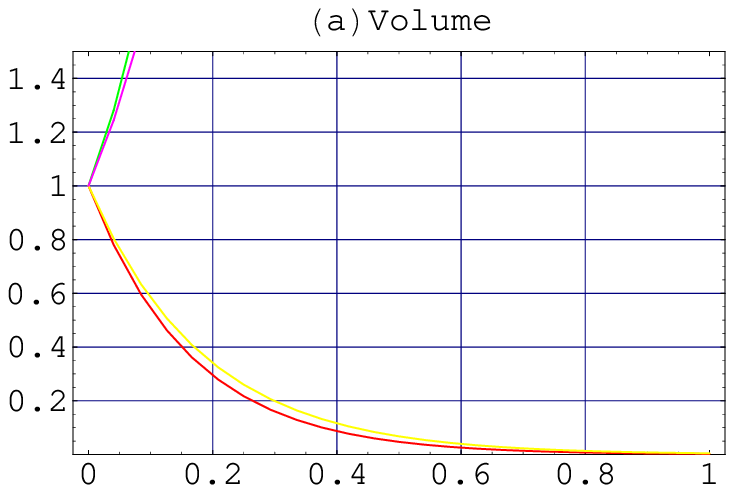,width=.45\linewidth}
\epsfig{file=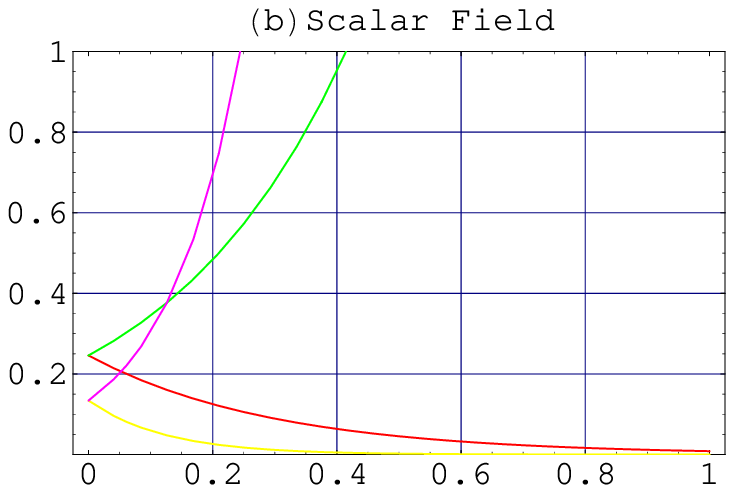,width=.45\linewidth} \caption{Plots (a) and
(b) represent the volume of the universe and scalar field versus
time $t$, respectively. Here red and green correspond to $(+,+)$ and
$(-,-)$ roots with $\omega=1.5$, respectively, while yellow and
purple lines indicate $(+,+)$ and $(-,-)$ roots with $\omega=1.2$,
respectively.}
\end{figure}
\begin{figure} \centering
\epsfig{file=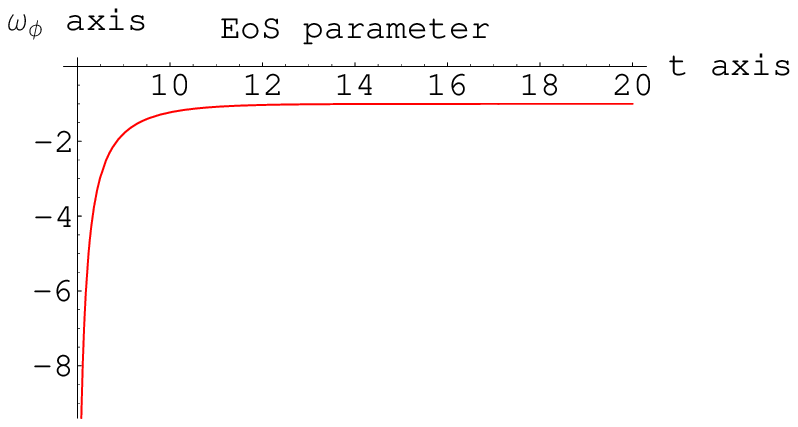,width=.45\linewidth}
\epsfig{file=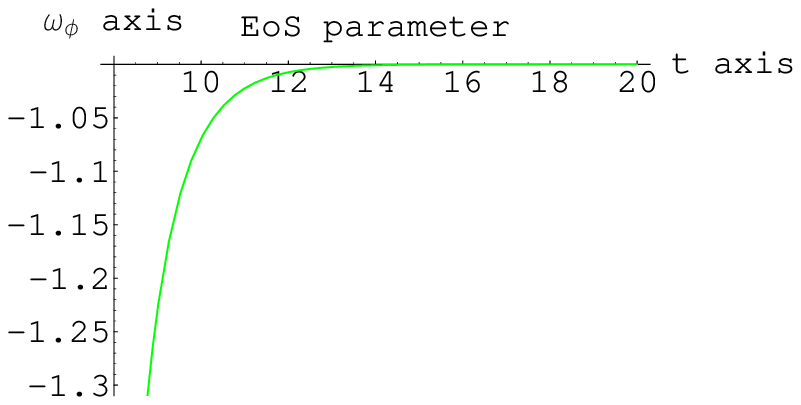,width=.45\linewidth} \caption{Plots (a) and
(b) show the EoS parameters for scalar field $\omega_\phi$, versus
time. Here red and green lines correspond to $(-,+)$ and $(+,-)$
roots with $\omega=200$, respectively.}
\end{figure}

For $(-,-)$ and $(+,+)$ roots, the scalar field is expanding, while
for $(+,-)$ and $(-,+)$, the scalar field is contracting with the
passage of time. This indicates that the scalar field plays a
dominant role in the later phase of cosmic expansion. The volume of
the universe is given by $V(t)=B^4=B_0^4\exp(4H_{2i}t)$ which shows
that the universe expands exponentially for different $H_{2i}$ roots
as shown in Figure \textbf{2(b)} (indicating infinite volume in
future). Moreover, the average scalar factor,
$a=B_0^{4/3}\exp(\frac{4H_{2i}t}{3})$, leads to negative value of
the deceleration parameter, which shows the rapid expansion in the
universe consistent with the observations. Figure \textbf{3}
represents the EoS parameter for scalar field
$\omega_\varphi=p_\varphi/\rho_\varphi$ versus time. These indicate
that the universe model lies in the phantom phase for all values of
time which is in agreement with the recent rapid expanding behavior
of the universe.

In the second case, we take $V_0=0$ (massless scalar field). Since
there exist two commutating symmetries $X_1$ and $X_2$, so we have
first-order DE $\frac{dF}{dH_2}=K(F,H_2)$ with one known symmetry.
Consequently, the integration in quadrature will be possible and
then Eqs.(\ref{12*}) and (\ref{13*}) yield
\begin{eqnarray}\nonumber
&&-\frac{dH_2}{H_2}=[(\frac{\varepsilon\omega}{6}
+\frac{m^2+2m+3}{2(1+2m)})+G^2\omega(\frac{(\varepsilon+6)\omega-3}{3(1+2m)})+\omega
G(\frac{\varepsilon(m+2)^2}{3(1+2m)}\\\nonumber
&&+\frac{3(m-1)}{3(1+2m)})][(\frac{2(\omega-1)+m(4\omega-1)}{(m+2)^2
-2\omega(1+2m)})\{(\frac{\varepsilon(m+2)^2
+6(m^2+2m+3)}{12(m+2)})\\\nonumber
&&+G^2(\frac{(m+2)(\varepsilon\omega-6)}{6(1+2m)}+\frac{\omega}{m+2}
+\frac{\omega(m+2)}{1+2m})\}+G\{\frac{2(\omega-1)+m(4\omega-1)}
{(m+2)^2-2\omega(1+2m)}\\\nonumber&&\times(\frac{\varepsilon(m+2)^2}{3(1+2m)}
-1+2\omega+\frac{m^2+2m+3}{1+2m})-(\frac{\varepsilon\omega}{6}
+\frac{m^2+2m+3}{2(1+2m)})\\\nonumber&&-G^2\omega(\frac{(\varepsilon+6)
\omega-3}{3(1+2m)})-\omega
G\frac{\varepsilon(m+2)^2+3(m-1)}{3(1+2m)}\}]^{-1}dG,
\end{eqnarray}
where $G=\frac{F}{H_2}$. The solution of this equation yields
directional Hubble parameter in terms of new parameter $G$. Since
the scalar field is massless, so BD coupling parameter should
satisfy the range $\omega\geq40,000$ as suggested by the solar
system experiments \cite{22}. For the present era, its solution can
be written as
\begin{equation}
H_2=H_{2,0}[(G-0.004916)^{0.000024}(G+19512.79)^{0.99995}(G-0.005085)^{0.0000256}],\\\label{**}
\end{equation}
where we have taken $m=0.5,~\omega=40,000$ and $H_{2,0}$ indicates
present value of the parameter $H_2$. From Eq.(\ref{12*}), the scale
factor becomes
\begin{eqnarray}\nonumber
-\frac{dB_2}{B_2}&=&\frac{2(\omega-1)+m(4\omega-1)}{2(1+2m)(m+2)}[(\frac{2(\omega-1)
+m(4\omega-1)}{(m+2)^2-2\omega(1+2m)})\{(\frac{\varepsilon(m+2)^2}{12(m+2)}
\\\nonumber
&+&\frac{6(m^2+2m+3)}{12(m+2)})+G^2(\frac{(m+2)(\varepsilon\omega-6)}{6(1+2m)}
+\frac{\omega}{m+2}+\frac{\omega(m+2)}{1+2m})\}\\\nonumber
&+&G\{(\frac{2(\omega-1)+m(4\omega-1)}{(m+2)^2-2\omega(1+2m)})(\frac{\varepsilon(m+2)^2}{3(1+2m)}
-1+2\omega+\frac{m^2+2m+3}{1+2m})
\\\nonumber&-&(\frac{\varepsilon\omega}{6}
+\frac{m^2+2m+3}{2(1+2m)})-G^2\omega(\frac{(\varepsilon+6)\omega-3}{3(1+2m)})-\omega
G\frac{\varepsilon(m+2)^2}{3(1+2m)}\\\nonumber&+&\frac{3(m-1)}{3(1+2m)}\}]^{-1}dG.
\end{eqnarray}

In the present era, for the choice $m=0.5$ and $\omega=40,000$, the
solution can be written as
\begin{equation}\nonumber
B=B_0[(G-0.004916)^{0.000024}(G+19512.79)^{0.99995}(G-0.005085)^{0.0000256}]^{-15999.8}.
\end{equation}
Further, the scalar field can be determined by the relationship
$\frac{d\varphi}{\varphi}=G\frac{da}{a}$, which follows from
$F=\frac{\dot{\varphi}}{\varphi}=GH$. The corresponding scalar field
takes the form
\begin{eqnarray}\nonumber
\varphi&=&\varphi_0\exp(-15999.8G)[(0.00508-G)^{1.3008\times10^{-7}}
(G+0.004916)^{-1.2008\times10^{-7}}\\\nonumber&\times&(G+19512.8)^{-19511.8}]^{-15999.8}.
\end{eqnarray}
The energy density is given by
\begin{eqnarray}\nonumber
\rho&=&\rho_0B_0^{-2.5}[(G-0.004916)^{0.000024}(G+19512.79)^{0.99995}(G\\\nonumber
&-&0.005085)^{0.0000256}]^{2.5\times15999.8}.
\end{eqnarray}
The symbols with 0 subscript indicate the present values. The time
related with the solution can be calculated by the expression
$t-t_0=\int \frac{da}{aH}$. For particular choice of parameters, it
becomes
\begin{eqnarray}\nonumber
t-t_0&=&\int[(0.800207(G+5.62517\times10^{-6})(G+0.975609))(\ln[(G\\\nonumber
&+&0.0035)^{0.000012}(G+26655.8)^{0.99997}
(G-0.0036)^{0.000012}])((G\\\nonumber
&-&0.00508)(G+0.004916)(G+19512.8))^{-1}]dG.
\end{eqnarray}
By inverting this expression, the parameter $G$ can be determined in
terms of time and hence the scale factor and the scalar field. The
above solutions are parametric solutions in new variable $G$, i.e.,
$a=a(G)$ and $\varphi=\varphi(G)$.

The above expression can be evaluated numerically and then by the
obtained set of data points, we can interpolate the function $G(t)$.
By adopting this procedure, we interpolate the function $G(t)$ using
polynomial interpolation and is given by
\begin{equation}
G(t)=63578.70834t^4-63719.54167t^3+22535.2479t^2-859.1271t+0.0059,\\\label{***}
\end{equation}
where we have used the initial condition $G(0)=0.0059$. Using this
value of $G$, all the expressions like energy density, scale
factors, scalar field and Hubble parameter can be discussed versus
time. It can be observed that the obtained model is not free from
singularities as the scalar field and scale factor become singular
for some particular values of $G$. Moreover, as $G\rightarrow0$, all
these quantities remain finite while as $G\rightarrow\infty$, only
the scalar field and the scale factor turn out to be zero. In this
case, the deceleration parameter is given by
$$q=-1-\frac{3}{m+2}(\frac{\dot{H}_2}{H_2^2}),$$
where $m=0.5$,while $H_2$ and $G$ are given by Eqs.(\ref{**}) and
(\ref{***}), respectively. Figure \textbf{4} shows that the
deceleration parameter remains negative for all values of time which
yields the accelerated expansion of the universe model and is
well-consistent with the recent observations.
\begin{figure} \centering
\epsfig{file=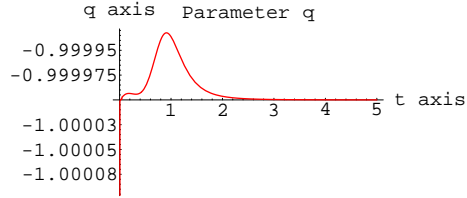,width=.45\linewidth} \caption{This shows the
deceleration parameter versus time.}
\end{figure}

\section{Summary}

The modified theories of gravity with action involving positive or
negative powers of curvature as an extra term can lead to a better
description of phenomenon of initial cosmic inflation and the
late-time cosmic acceleration. The main objective of this paper is
to evaluate the Noether and Noether gauge symmetries for some
homogeneous universe models in the framework of non-vacuum
scalar-tensor gravity with inverse curvature correction term, i.e.,
$R^{-1}$. For this purpose, we have applied the Noether gauge
symmetry analysis to non-flat FRW universe model. Furthermore, we
have discussed the Noether and Noether gauge symmetries for BI
universe model. In both cases, the matter part of Lagrangian has
been taken as perfect fluid. We have constructed the field potential
and the coupling function by requiring the existence of Noether
symmetries.

In the case of FRW universe model, we have a system of 11 PDEs for
Noether gauge symmetries of the constructed point like Lagrangian.
In literature \cite{8}, the Noether symmetries of the same
Lagrangian in vacuum has been discussed, where the model with dust
matter seem to be more physical, but the Noether symmetries cannot
always exist. We have extended this work by exploring more general
symmetries, i.e., Noether gauge symmetries by introducing perfect
fluid matter part in the Lagrangian. We have found that the Noether
symmetry generator exists with non-zero gauge function in matter
dominated phase. It is seen that the gauge term turns out to be a
dynamical quantity and the integral of motion associated with the
dynamics of Lagrangian exist. We have also specified the form of BD
coupling function and the field potential. In this case, the BD
coupling function turns out to be a constant quantity, while the
field potential is given by the power law form.

Next, it is shown that the Noether as well as Noether gauge
symmetries exist for the flat BI universe model with perfect fluid
using point like Lagrangian with curvature corrected term. The
existence of Noether symmetry generator allows zero coupling
function and quadratic potential with zero integral of motion. The
Noether gauge symmetry generators yield the constant gauge function
with quadratic potential and variable BD parameter,
$\omega=c_4/\varphi^{2p}$. The behavior of BD coupling parameter is
dependent on the parameter $p$. We have also determined the
respective conserved quantities in this case.

Finally, we have evaluated the BI solutions using scaling or
dilatational symmetries. Since it is difficult to find the BI
cosmological solutions in the curvature corrected configuration, so
we take $\mu=0$ in the action, i.e., zero curvature correction term
and constant BD coupling parameter. For this purpose, two cases have
been taken into account. In the first case, it is seen that both the
scale factor and the scalar field evolve exponentially yielding
deceleration parameter $q=-1$ which is compatible with the
observations and inflationary scenario. Furthermore, the EoS
parameter for scalar field turns out to be negative
$\omega_\varphi<-1$ for $(-,+)$ and $(+,-)$ roots only that confirms
the accelerating phase of the universe model. The graphs of scale
factor and energy density have also been given. In the second case,
there exist two symmetry generators and consequently, the
integration in quadrature is possible. By introducing a new
parameter $G=F/H_2$, the forms of scale factor, scalar field,
directional Hubble parameters and energy density have been
calculated. It is observed that the obtained solution is parametric
in the variable $G$. The relation of new variable $G$ in terms of
time has also been given. By solving the function $G$ numerically
using polynomial interpolation, all the cosmological parameters can
be discussed versus time. In this respect, the plot of the
deceleration parameter versus time has been given which shows that
the parameter takes negative values for all values of time. This is
well-consistent with the current rapid expanding behavior of the
universe. It would be interesting to discuss the cylindrically or
plane symmetric models using Noether symmetry analysis in the
framework of scalar-tensor gravity with curvature correction.

\renewcommand{\theequation}{A\arabic{equation}}
\setcounter{equation}{0}

\end{document}